\documentclass[amsfonts,amsmath,prd,preprint,nofootinbib]{revtex4}
\newcommand{\beq}{\begin{equation}}
\newcommand{\eeq}{\end{equation}}

\newcommand{\bsp}{\begin{split}}

\usepackage{epsfig,bbm,cancel,ulem}
\usepackage[breaklinks=true]{hyperref}
\usepackage{amsmath}
\usepackage{slashed}
\usepackage{latexsym}
\usepackage{braket}
\graphicspath{{./figures/}}

\begin{document}

\title{Coleman-de Luccia Tunneling Wave Function}
\author{J.~Kristiano, R.~D.~Lambaga, and H.~S.~Ramadhan}
\affiliation{Departemen Fisika, FMIPA, Universitas Indonesia, Depok, 16424, Indonesia. }
\def\changenote#1{\footnote{\bf #1}}


\begin{abstract}

We revisit the famous Coleman-de Luccia formalism for decay of false vacuum in gravitational theory. Since the corresponding wave function is time-independent we argue that its instanton's interpretation as the decay rate probability is problematic. We instead propose that such phenomenon can better be described by the Wheeler-de Witt's wave function. To do so, the Hamilton-Jacobi formalism is employed in the WKB approximation. The scalar and gravitational fields can then be treated as a two-dimensional effective metric. For a particular case of dS-to-dS tunneling, we calculated the wave function and found that it depends only on the potential of the false vacuum. In general, this alternative approach might have significant impact on the study of very early universe and quantum cosmology.
 
\end{abstract}

\maketitle

\section{Introduction} \label{sec:introduction}

Decay of metastable state was explained by Coleman and Callan (CC) using the Euclidean action formalism~\cite{coleman:1977PRD,callan:1977PRD}. Consider a system with two homogeneous stable equilibrium states with different potential levels. Metastable state is the state in the local minimum of potential, commonly called the {\it false vacuum}. This state might decay to the {\it true vacuum}, the state of global minimum of the potential, by quantum tunneling. Coleman and Callan derived the decay rate using Euclidean path integral to find the imaginary part of energy. The decay rate per unit volume can be expressed as $\Gamma/\mathcal{V} = Ae^{-B}$, where $B$ is Euclidean action of classical solution and $A$ is the coefficient found by quantum correction. They derived the decay rate in quantum mechanics formalism, but it can be extended into quantum field theory. The starting point is the Euclidean path integral
\begin{equation}
Z = \bra{x_f} e^{-HT} \ket{x_i} = N \int_{x(0) = x_i}^{x(T) = x_f} [dx] e^{-S_E[x]},
\label{pathintegral}
\end{equation}
where $H$ is the Hamiltonian, $x_i$ and $x_f$ are the initial and final points respectively, and $T$ is Euclidean time interval. By inserting the left-hand side of Eq.~\eqref{pathintegral} with a complete set of Hamiltonian eigenstates $H \ket{n} = E_n \ket{n}$, the equation becomes
\begin{equation}
Z = \sum_{n} e^{-E_n T} \phi_n(x_i) \phi_n^*(x_f),
\end{equation}
where $\phi_n(x_i) = \left\langle n | x_i \right\rangle$ and $\phi_n^*(x_f) = \left\langle x_f | n \right\rangle$ are wave function of initial and final state respectively. For $T\rightarrow\infty$, the lowest energy term dominates the path integral. Then, the lowest energy can be expressed in the form
\begin{equation}
E_0 = - \lim_{T \rightarrow \infty} \frac{\ln Z}{T},
\end{equation}
and the corresponding decay rate is function of imaginary part of $E_0$. Consider the time dependent state
\begin{equation}
\ket{\Psi(t)} = \sum_{n} c_n e^{-iE_n t} \ket{n},
\end{equation}
where $E_n$ is a complex variable, such as $E_n = E^{(R)}_n - i E^{(I)}_n$ and $E^{(R)}_n, E^{(I)}_n \in \mathbb{R}$. By determining the coefficient $c_n$ from the initial state $\ket{\Psi(0)}$, the time dependent state can be expressed as
\begin{equation}
\ket{\Psi(t)} = \sum_{n} \left\langle n | \Psi(0) \right\rangle ~e^{-E^{(I)}_n t} e^{-i E^{(R)}_n t} \ket{n}.
\end{equation}
After a long time, the ground state term $n=0$ will dominate the summation and become
\begin{equation}
\ket{\Psi(t)} \approx \left\langle 0 | \Psi(0) \right\rangle e^{-E^{(I)}_0 t} e^{-i E^{(R)}_0 t} \ket{0}.
\end{equation}
Projection of the state to position basis is
\begin{equation}
\Psi(x,t) \approx \left\langle 0 | \Psi(0) \right\rangle e^{-E^{(I)}_0 t} e^{-i E^{(R)}_0 t} \phi_0(x),
\end{equation}
and the corresponding probability to find the particle in false vacuum $x_{FV}$ is
\begin{equation}
P(x_{FV},t) = \Psi^*(x_{FV},t) \Psi(x_{FV},t) = \lvert \left\langle 0 | \Psi(0) \right\rangle \phi_0(x_{FV}) \rvert^2  e^{-2E^{(I)}_0 t},
\end{equation}
from which we can deduce the decay rate of probability to find the particle in false vacuum is $\Gamma = 2E^{(I)}_0 = -2 ~\mathrm{Im} (E_0)$.

Prior study to the barrier penetration in many dimensions was done by Banks, Bender, and Wu (BBW)~\cite{banks:1973PRD}. The tunneling amplitude for a particle with mass $m$ and energy $E$ is given by the WKB formula
\begin{equation}
T = \exp\left(-\int ds \sqrt{2m(V-E)}\right),
\end{equation}
with $ds$ is the path which minimizes the integral above, and thus satisfies
\begin{equation}
\delta \int ds ~\sqrt{V-E} = 0.
\end{equation}

BBW made a formalism for tunneling wave function, while CC formulate how to calculate the decay rate. Tunneling wave function is the ratio between time-independent wave function at one side of barrier to the other side with the same potential level, while the decay rate is the ratio between probability of finding the state in the same region after certain time. In the literature the relation between the tunneling probability and the decay rate is often skipped, mostly by assuming that they are equivalent. A closer look, however, shows that there is a subtle difference between them; the former {\it can} have no time dependence. Thus it is not perfectly clear how to draw their equivalence. Recently, such relation was discussed in great detail by Andreassen et.al \cite{andreassen:2017PRD}. Consider a state initially in the false vacuum. The probability of finding the state in the same region after time $T$ is
\begin{equation}
P_{FV}(T) = \int_{FV} dx |\psi(x,T)|^2 \sim \exp(-\Gamma T),
\end{equation}
where it is expected to fall exponentially. So, the decay rate is
\begin{equation}
\Gamma = - \frac{1}{P_{FV}} \frac{dP_{FV}}{dT}.
\end{equation}
We recall the result obtained by Andreassen et.al \cite{andreassen:2017PRD}. Consider a one dimensional barrier penetration problem, with false vacuum located in $x=a$ and the same potential level at the right side of barrier located in $x=b$. The relation between the decay rate and the tunneling wave function is
\begin{equation}
\Gamma = \frac{p_b}{m} \frac{|\phi_E(b)|^2}{\int_0^b dx |\phi_E(x)|^2},
\end{equation}
where $\phi_E(x)$ is the wave function,
\begin{equation}
\phi_E(x) = \frac{A}{\left( 2m\left[E-V(x)\right] \right)^{1/4}} \exp{\left(-\int_a^x dy \sqrt{2m[V(y)-E]}\right)},
\end{equation}
the coefficient $A$ is an $x$-independent normalization constant, and
\begin{equation}
p_b = -\frac{i}{2} \left[ \frac{\phi_E^* \partial_x \phi_E - \phi_E \partial_x \phi_E^*}{|\phi_E|^2} \right]_{x=b}.
\end{equation}

When gravity is added, the inequivalence discussed above is more severe, precisely because the notion of decay rate is, strictly-speaking, ill-defined. The effects of gravitation to the decay rate was first studied by Coleman and de Luccia (CdL) \cite{cdluccia:1980PRD}. The action consists of scalar and gravitational fields, with homogeneous and isotropic Euclidean metric. Contrary to naive expectation, they showed that the effects of gravitation are not negligible but have critical contribution to the decay rate instead. Unfortunately, the use of Euclidean action formalism to describe the decay rate or tunneling probability of fields with gravitation has several problems. Vilenkin tried to interpret the Euclidean action as tunneling probability for creation of the universe, but it leads to positive exponential probability \cite{vilenkin:1983PRD}. Later he used Wheeler-DeWitt tunneling wave function and it leads to negative exponential probability \cite{vilenkin:1984PRD}. Another problem in Euclidean action formalism was explained by Feldbrugge, Lehners, and Turok (FLT) \cite{feldbrugge:2017PRD}. They argue that the Lorentzian path integral is a better starting point for quantum cosmology. They found that the problem in Euclidean approach is that the inclusion of topologically-nontrivial manifold renders the Euclidean action unbounded below. 

The main problem in Euclidean approach for gravitational theory is that the corresponding Hamiltonian becomes the constraint for the wave function, $H \Psi = 0$, which is nothing but the Wheeler-de Witt (WdW) equation \cite{dewitt:1967PR}. The consequence is that the wave function is time-independent. This makes the interpretation of the exponential of Euclidean action as a decay rate problematic. Mathematically, the problem is as follows: the path integral over compact Euclidean four-geometry and scalar field is given by
\begin{equation}
\label{Z}
Z = \bra{g'_{\mu \nu},\phi'} e^{-HT} \ket{g_{\mu \nu}, \phi} = N \int [dg_{\mu \nu}] [d\phi] e^{-S_E[g_{\mu \nu}, \phi]}.
\end{equation}
Since $H\Psi=0$ we cannot insert a complete set of Hamiltonian eigenstates in the left-hand side of Eq.~\eqref{Z}. We thus cannot get the decay rate from the ground state energy, because the Hamiltonian constraint makes the eigenvalue always equal to zero. We argue that the tunneling wave function is more suitable to describe CdL effects, rather than the Euclidean action.

There is an alternative explanation for the notion of decay rate in gravitational theory. Mithani and Vilenkin \cite{mithani:2015PRD} introduce semiclassical superspace variables, which play the role of clock. They calculated decay rate in the "simple harmonic universe" (SHU) model. The role of clock was played by a homogenous, massless, minimally coupled scalar field. Another alternative was pointed out by Feng and Matzner \cite{feng:2017PRD}. They found that variations of metric at spatial boundary can be used to describe time evolution. The role of time in gravitational theory is governed by boundary conditions on the gravitational field at the spatial boundary. They called the theory an extended WdW equation. The concept of decay rate in gravitational theory may be developed along this line.

\section{The WdW Equation for CdL Tunneling}\label{sec:wdw}

Emphasizing the difference between the notion of {\it the tunneling probability} and {\it the decay rate}, in this paper we study the CdL phenomenon from the point of view of (Lorentzian) wave function. To begin, let us start from the CdL action \cite{cdluccia:1980PRD}
\begin{equation}
S = \int d^4x \sqrt{-g} \left[ -\frac{1}{2}g^{\mu \nu} \partial_\mu \phi \partial_\nu \phi - V(\phi) + \frac{R}{2\kappa} \right],
\end{equation}
where $\phi$ and $V(\phi)$ are the scalar field and its potential, $g_{\mu \nu}$ and $g$ is metric field and its determinant, $R$ is Ricci scalar, and $\kappa = 8\pi G$.  Consider homogeneous and isotropic metric and scalar fields,
\begin{eqnarray}
ds^2&=&-dt^2 + a^2(t) [d\chi^2 + \sin^2\chi (d\theta^2 + \sin^2\theta d\phi^2)],\nonumber\\ 
\phi&=&\phi(t).
\end{eqnarray} 
With those ansatz, the action becomes
\begin{equation}
S = \int dt ~2\pi^2 a^3 \left[ \frac{1}{2} \dot{\phi}^2 - V(\phi) + \frac{R}{2\kappa} \right],
\end{equation}
and the corresponding Lagrangian is
\begin{equation}
L = 2\pi^2 a^3 \left[ \frac{1}{2} \dot{\phi}^2 - V(\phi) + \frac{R}{2\kappa} \right],
\end{equation}
with Euler-Lagrange equation
\begin{eqnarray}
\label{EL}
\dot{a}^2 + 1 - \frac{1}{3}\kappa a^2 \left[ V(\phi) + \frac{1}{2} \dot{\phi}^2 \right] = 0, \nonumber\\
\ddot{\phi} + 3\frac{\dot{a}}{a} \dot{\phi} + \frac{dV}{d\phi} = 0.
\end{eqnarray}
which is also the tt-component of the Einstein equation.
The canonical momentum of both field is
\begin{equation}
\pi_\phi = \frac{\partial L}{\partial \dot{\phi}} = 2\pi^2 a^3 \dot{\phi},
\end{equation}
\begin{equation}
\pi_a = \frac{\partial L}{\partial \dot{a}} = -\frac{12\pi^2}{\kappa} a \dot{a}.
\end{equation}

The Hamiltonian is found by means of the Legendre transformation
\begin{eqnarray}
H &=& \pi_\phi \dot{\phi} + \pi_a \dot{a} - L \nonumber\\
&=& -\frac{6\pi^2}{\kappa}a \left\lbrace \dot{a}^2 + 1 - \frac{1}{3}\kappa a^2 \left[ \frac{1}{2}\dot{\phi}^2 + V(\phi) \right]  \right\rbrace,
\end{eqnarray}
which is equal to zero because the terms in the bracket exactly satisfy Eq.~\eqref{EL}. The Hamiltonian becomes a constraint for the wave function, $H \Psi = 0$. Expressing $H$ in terms of canonical momentum and field, and change the momentum to field operator 
\begin{eqnarray}
\pi_\phi = -i \frac{\partial}{\partial \phi}, \nonumber\\
\pi_a = -i \frac{\partial}{\partial a},
\end{eqnarray}
the constraint becomes
\begin{equation}
\label{Hconstraint}
H \Psi = \left\lbrace -\frac{1}{4\pi^2 a^3} \frac{\partial^2}{\partial \phi^2} + \frac{\kappa}{24\pi^2 a} \frac{\partial^2}{\partial a^2} - \frac{6\pi^2}{\kappa} a + 2\pi^2 a^3 V(\phi) \right\rbrace \Psi = 0,
\end{equation}
which is the WdW equation. We can see that gravity modifies the potential of the wave function. It now consists not only of the scalar field potential $V(\phi)$, but also there is contribution from the scale factor $a$. This potential makes the wave function is not separable in terms of $a$ and $\phi$.

\section{The Hamilton-Jacobi of CdL Tunneling}\label{sec:hj}

The general solution to the WdW equation can be expressed in the form $\Psi(a,\phi) = A(a,\phi) ~e^{-i\Phi(a,\phi)}$. Substituting it into Eq.~\eqref{Hconstraint} and employing the WKB-approximation (neglecting the second derivative of $A$), we get the following differential equation
\begin{equation}
-\frac{6}{\kappa a^2} \left( \frac{\partial \Phi}{\partial \phi} \right)^2 + \left( \frac{\partial \Phi}{\partial a} \right)^2 + 2U(\phi,a) = 0,
\label{dewittwkb}
\end{equation}
where $U(\phi,a)$ is the {\it effective} potential,
\begin{equation}
2U(\phi,a)\equiv\left( \frac{12\pi^2}{\kappa} \right)^2 a^2 \left\lbrace 1 - \frac{1}{3}\kappa a^2 V(\phi) \right\rbrace. 
\end{equation}

In quantum mechanics, the differential equations of the WKB-approximation correspond to the Hamilton-Jacobi (HJ) equation in classical mechanics. Therefore, we may call~\eqref{dewittwkb} the Hamilton-Jacobi equation with two degrees of freedom (scale factor $a$ and scalar field $\phi$). The general form of the HJ equation is
\begin{equation}
(\nabla \Phi) \cdot (\nabla \Phi) + 2U(\phi,a) = 0.
\end{equation}
The first term can be expressed in the form of $g^{ij} \partial_i \Phi \partial_j \Phi$. The scale factor and the scalar field form a new coordinate system $q^i = (\phi , a)$. Matching the general form of Hamilton-Jacobi equation to equation (\eqref{dewittwkb}), we get the ``internal" metric
\begin{equation}
g^{ij} = \mathrm{diag} \left( -\frac{6}{\kappa a^2} , 1 \right),
\end{equation}
and the ``internal" line element
\begin{equation}
ds^2 = g^{ij} dq_i dq_j = -\frac{1}{6}\kappa a^2 d\phi^2 + da^2.
\end{equation}
The interesting thing is that the metric component $g^{\phi \phi}$ has opposite sign to $g^{aa}$. Therefore, the line element $ds$ is not always a real, {\it i.e.,} it can be imaginary. This makes an important consequence to the tunnelling wave function, which we discuss later. The solution of the HJ equation is
\begin{eqnarray}
\label{Phi}
\Phi &=& \int ds ~[-2U(\phi,a)]^{1/2} \nonumber\\
&=& \int \left[ -\frac{1}{6} \kappa a^2 d\phi^2 + da^2 \right]^{1/2}[-2U(\phi,a)]^{1/2},
\end{eqnarray}
with $\phi(t)$ and $a(t)$ are the solutions that makes $\Phi$ stationary, $\delta \Phi = 0$. To simplify the problem, we take $a$ as an independent variable so we can express $\phi=\phi(a)$, and then define $\alpha\equiv a\sqrt{\kappa/3}$. The differential equation satisfied by $\phi(\alpha)$ is then found by variational calculus
\begin{equation}
\phi'' = \left\lbrace \frac{3}{\kappa} \partial_\phi V - \frac{\phi'}{\alpha} (1 - 2\alpha^2 V) \right\rbrace \left( \frac{1 - \frac{1}{6}\kappa \alpha^2 \phi'^2}{1 - \alpha^2 V} \right) - \frac{\phi'}{\alpha} \left( 2 - \frac{1}{6} \kappa \alpha^2 \phi'^2 \right),
\label{dewitthj}
\end{equation}
where prime denotes derivative with respect to $\alpha$.

\begin{figure}[htbp]
	\centering\leavevmode
	\epsfysize=7.5cm
	\epsfbox{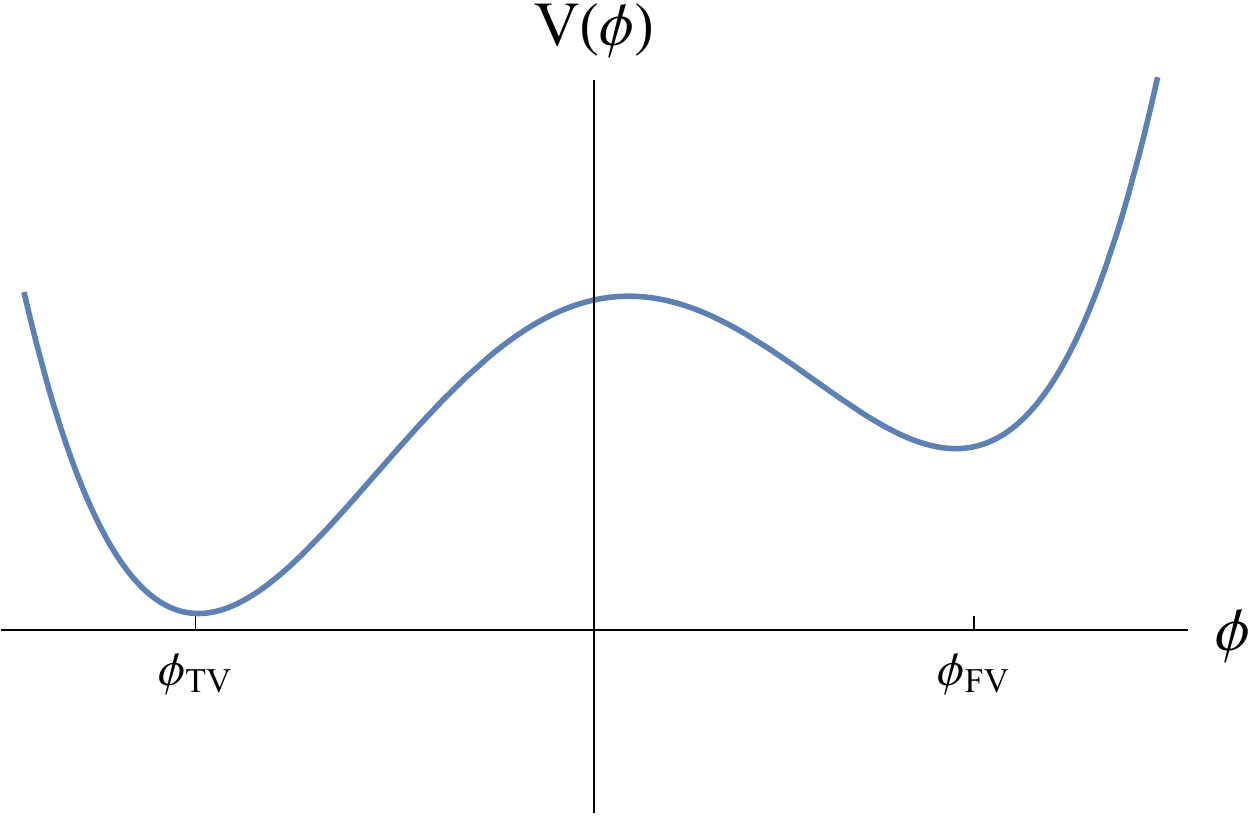}
	\caption{Plot of $V(\phi)$.}
	\label{fig:V}
\end{figure}

\begin{figure}[htbp]
	\centering\leavevmode
	\epsfysize=7.5cm
	\epsfbox{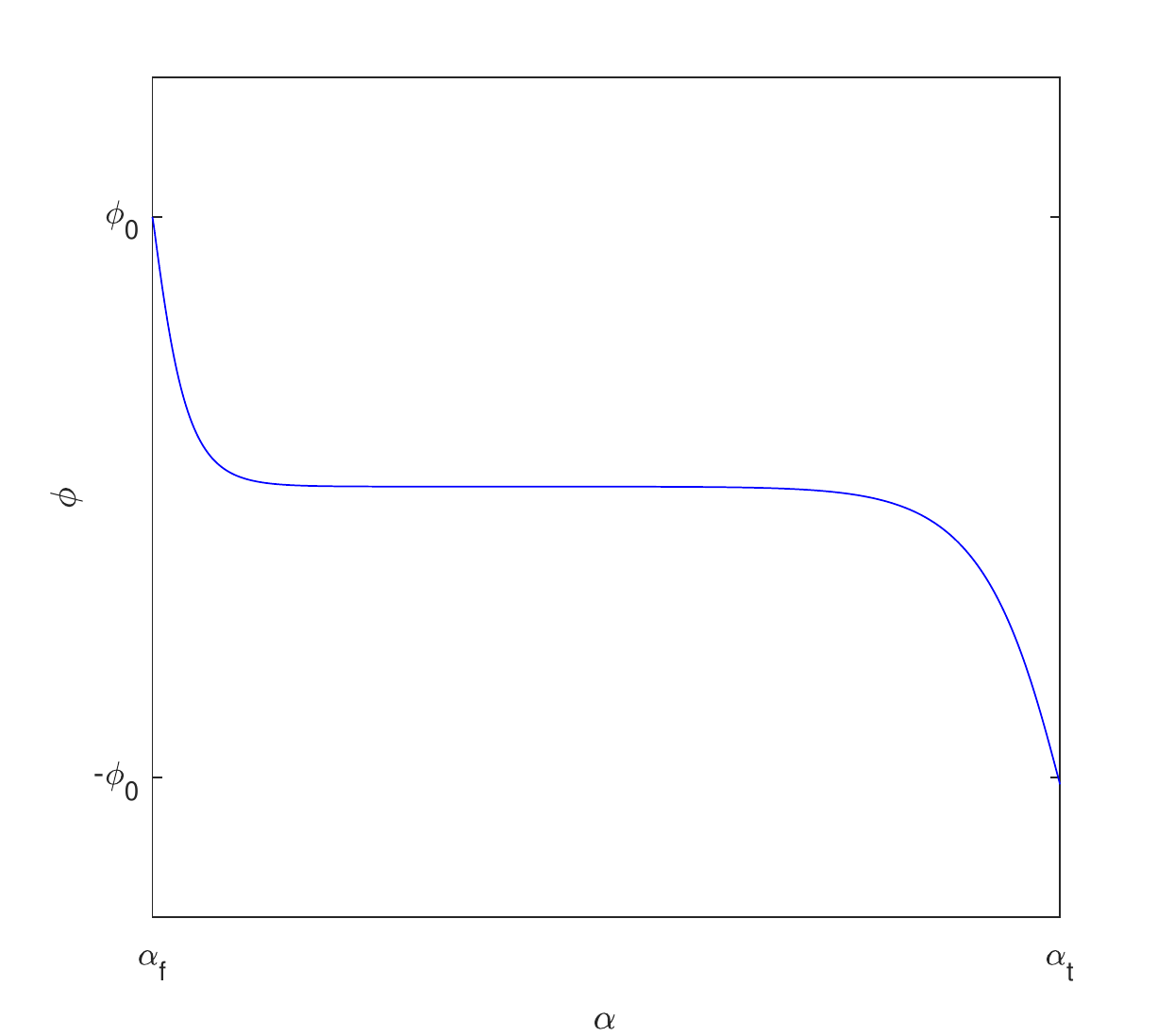}
	\caption{Profile of $\phi(\alpha)$.}
	\label{fig:HJ}
\end{figure}

\begin{figure}[htbp]
	\centering\leavevmode
	\epsfysize=7.5cm
	\epsfbox{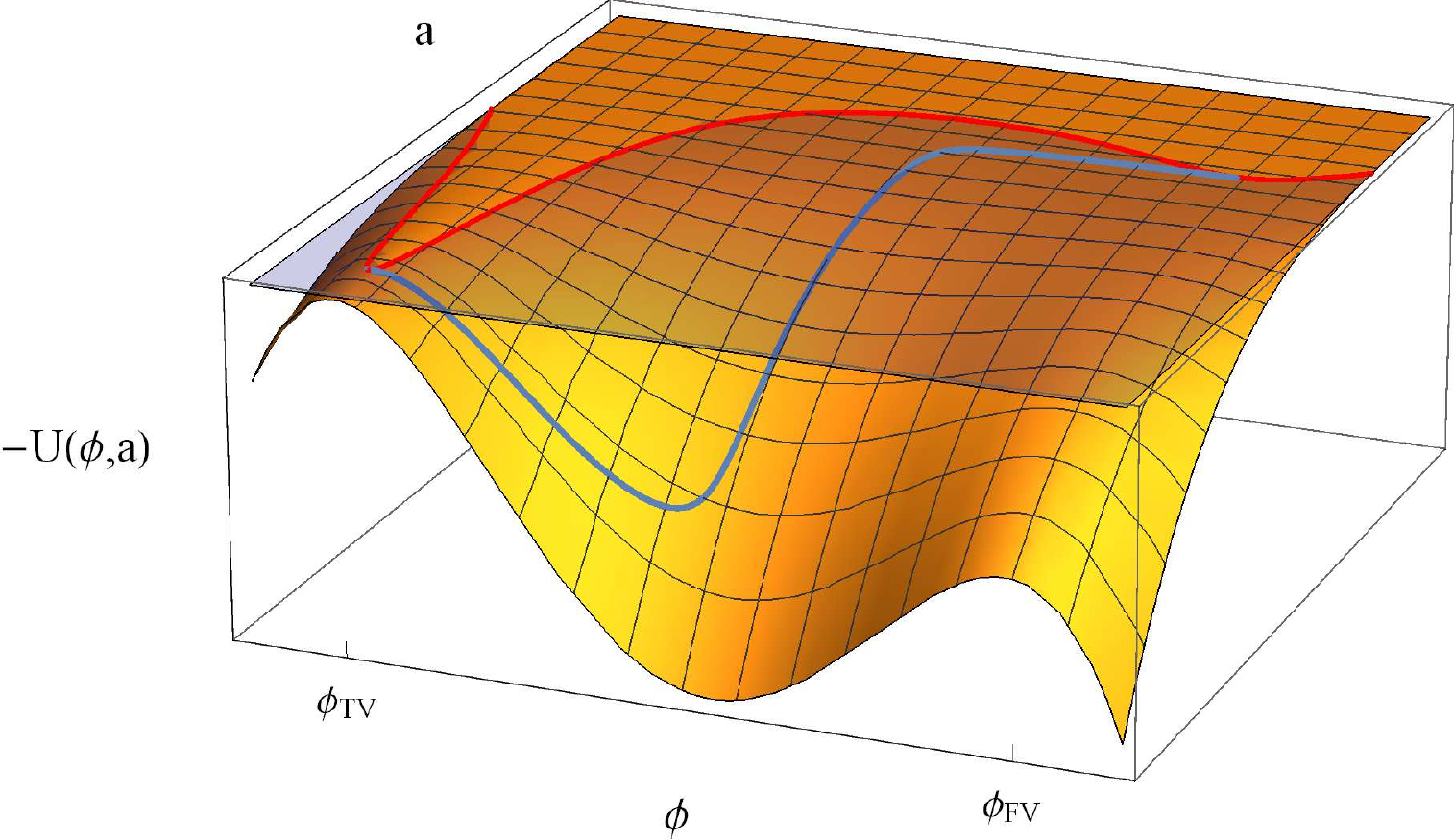}
	\caption{Plot of $-U(\phi,a)$. The blue line indicated a rough approximation of the profile of $\phi(a)$ given by Fig.~\ref{fig:HJ}. The red line indicated $-U(\phi,a) =0$.}
	 \label{fig:U}
\end{figure}

To solve~\eqref{dewitthj} we need a specific $V(\phi)$ function. Consider the following scalar field potential
\begin{equation}
V(\phi) = \frac{\lambda}{8} (\phi^2 - \phi_0^2)^2 + (\epsilon_f - \epsilon_t) \left( \frac{\phi + \phi_0}{2\phi_0} \right) + \epsilon_t,
\end{equation}
with assumption $\epsilon_f, \epsilon_t \ll \lambda$. This potential has a local minimum at $V(\phi_0) = \epsilon_f$ and a global minimum at $V(-\phi_0) = \epsilon_t$. The shape of $V(\phi)$ can be seen in Fig.~\ref{fig:V}. 
The boundary condition for \eqref{dewitthj} is the classical turning point $U(\phi,a) = 0$. Consider tunneling from false vacuum to true vacuum, the initial point is $(\phi_0, 1/\sqrt{\epsilon_f})$ and the final point is $(-\phi_0, 1/\sqrt{\epsilon_t})$. Unfortunately, the boundary condition makes the differential equation singular because the factor $1 - \alpha^2 V$ in the denominator is equal to zero.The numerator, then, must be zero; i.e $\phi'$ must have fixed value $\phi'(\alpha_f) = \pm \sqrt{\frac{6}{\kappa \alpha_f^2}}$ and $\phi'(\alpha_t) = \pm \sqrt{\frac{6}{\kappa \alpha_t^2}}$ at initial and final point respectively. It is futile to solve~\eqref{dewitthj} analytically. Therefore we resort to numerical solution. A typical profile of $\phi(\alpha)$ is presented in Fig.~\ref{fig:HJ}. Here we set $\sqrt{6/\kappa} \gg \phi_0$ in order not to violate the Grand Unified Theory (GUT) scenario. In natural units, $\sqrt{6/\kappa} \approx 10^{18} ~\mathrm{GeV}$ while $\phi_0 \approx 10^{14} ~\mathrm{GeV}$ \cite{albrecht:1982PRL}. We also embed this profile to $-U(\phi,a)$ in Fig.~\ref{fig:U}.
At first the plot looks like an {\it inverted kink}. However, there is a difference between the two. Our solution is closed (or compact), while the kink solution is open, {\i.e.,} extending to infinity by asymptotically approaching the vacuum.

To calculate the integral~\eqref{Phi} we can just integrate, by brute force, the numerical solution of $\phi(\alpha)$. However, it is more instructive to approximate the profile solution as three separate regions. The first is a straight line with gradient $\phi' = - \sqrt{\frac{6}{\kappa \alpha_f^2}}$ from $\phi = \phi_0$ to $\phi = 0$. The second is a horizontal line at $\phi = 0$. The third region is again a straight line with gradient $\phi' = - \sqrt{\frac{6}{\kappa \alpha_t^2}}$ from $\phi = 0$ to $\phi = -\phi_0$. We believe this is a reliable approximation since we accommodate the characteristic of the differential equation, which has fixed value of gradient at initial and final point. Interestingly, as we shall see, the tunneling contribution comes only from the first region. The relevant contribution to probability comes from the imaginary part of $\Phi$, which in our case is obtained only from the first region. Thus, contrary to the usual quantum mechanics problem the tunneling contribution comes from the path $ds$, not from the potential part, because the line element $ds$ can be imaginary. Substituting the solution $\phi(\alpha)$ and its gradient to $\Phi$, we obtain
\begin{equation}
\Phi = i \frac{36\pi^2}{\kappa^2} \left( \frac{\kappa}{6} \right)^{1/2} \int_{\phi_0}^0 d\phi ~[\alpha^2 - \alpha_f^2]^{1/2} ~\alpha [\alpha^2 V - 1]^{1/2}, 
\end{equation}
with $\alpha(\phi)$ is a straight line function
\begin{equation}
\alpha = \alpha_f \left\lbrace 1 + \left( \frac{\kappa}{6} \right)^{1/2} (\phi_0 - \phi) \right\rbrace.
\end{equation}
We may neglect $1$ compared to $\alpha^2 V$ because $\alpha^2 V \approx V/\epsilon_f \gg 1$, so we get
\begin{eqnarray}
\Phi &=& i \frac{36 \pi^2}{\kappa^2} \left( \frac{\kappa}{6} \right)^{1/2} \left( \frac{\lambda}{8} \right)^{1/2} \alpha_f^3 \nonumber\\
&& \int_{\phi_0}^0 d\phi ~(\phi_0^2 -\phi^2) \left[ 1 + \left( \frac{\kappa}{6} \right)^{1/2} (\phi_0 - \phi) \right]^2 \sqrt{ 2 \left( \frac{\kappa}{6} \right)^{1/2} (\phi_0 - \phi) + \frac{\kappa}{6} (\phi_0 - \phi)^2 }.
\end{eqnarray}
Expanding it in power of $\kappa$
\begin{equation}
\Phi \approx -i \left( \frac{36 \sqrt{2} \pi^2}{\kappa^2} \right) \left( \frac{\kappa}{6\epsilon_f^2} \right)^{3/4} \left( \frac{\lambda}{8} \right)^{1/2} \left( \frac{18}{35} \phi_0^{7/2} \right).
\end{equation}
The tunneling probability is proportional to $P \sim \exp(-2 |\Phi|)$. The non-zero contribution comes when the fields go from the false vacuum to the local maximum. After that, the fields stuck in the local maximum and ``roll down" to the global minimum. The value of the global minimum does not contribute to the wave function. 

Using the usual GUT parameters \cite{albrecht:1982PRL}, $\phi_0 = (1 \times 10^{14}) ~\mathrm{GeV}$, $\lambda = 80$, $\epsilon_f = (4 \times 10^{56}) ~\mathrm{GeV}^4$, $\epsilon_t = (1 \times 10^{56}) ~\mathrm{GeV}^4$, and fundamental constant $\kappa = (1.7 \times 10^{-37}) ~\mathrm{GeV}^{-2}$, the tunneling probability is proportional to $P \sim \exp(-5.0 \times 10^{12})$, such a very small probability. For comparison, the ordinary CdL decay rate has order of magnitude $\Gamma/\mathcal{V} \sim \exp(-6.2 \times 10^3)$.

\section{Conclusions}\label{sec:conc}

In conclusion, we wish to draw the reader's attention to the subtle distinction between the tunneling probability and the decay rate. While the two are perfectly well-defined and proportionally-related in non-gravitational theories, the story is not that trivial when gravity is included. Taking the Wheeler-de Witt (WdW) into account it can be shown that the notion of decay rate becomes ill-defined; thus the Coleman-de Luccia (CdL) phenomenon is better perceived as a tunneling, rather than decay, phenomenon. We use the BBW formalism to compute the CdL wave function. We consider a homogeneous and isotropic ansatz to the gravitational and scalar field. We find that the ``{\it compact rotated kink}" is the solution of Hamilton-Jacobi differential equation, which fits in the GUT scenario. This wave function only depends on the potential of false vacuum. As in the original CdL bounce, gravity still has non-perturbative effect on the tunneling probability. It is that now the whole picture is described in a (relatively) more consistent quantum-gravitational formalism. Our numerical probability value is reported to be much smaller than the original CdL. From realistic cosmological point of view this is a good news.

\section{Acknowledgement}

We thank Alex Vilenkin for the enlightening discussions and useful comments on the original manuscript. This work is partially funded by the Q1Q2-Grants from Universitas Indonesia under the contract No.~NKB-0277/UN2.R3.1/HKP.05.00/2019 (JK) and NKB-0270/UN2.R3.1/HKP.05.00/2019 (RDL and HSR).


\begin{thebibliography}{99}

\bibitem{coleman:1977PRD} 
  S.~Coleman, 
  Phys.\ Rev.\ D {\bf 15}, 2929; {\bf 16}, 1248(E) (1977).
\bibitem{callan:1977PRD} 
  C.~G.~Callan and S.~Coleman,
  Phys.\ Rev.\ D {\bf 16}, 1762 (1977).
\bibitem{banks:1973PRD} 
  T.~Banks, C.~M.~Bender, and T.~T.~Wu,
  Phys.\ Rev.\ D {\bf 8}, 3346 (1973).
\bibitem{andreassen:2017PRD} 
  A.~Andreassen, D.~Farhi, W.~Frost and M.~D.~Schwartz,
  Phys.\ Rev.\ D {\bf 95}, 085011 (2017).
\bibitem{cdluccia:1980PRD} 
  S.~R.~Coleman and F.~De Luccia,
  Phys.\ Rev.\ D {\bf 21}, 3305 (1980).
\bibitem{vilenkin:1983PRD} 
  A.~Vilenkin, 
  Phys.\ Rev.\ D {\bf 27}, 2848 (1983).
\bibitem{vilenkin:1984PRD} 
  A.~Vilenkin, 
  Phys.\ Rev.\ D {\bf 30}, 509 (1984).
\bibitem{feldbrugge:2017PRD} 
  J.~Feldbrugge, J.~L.~Lehners and N.~Turok,
  Phys.\ Rev.\ D {\bf 95}, 103508 (2017).
\bibitem{dewitt:1967PR} 
  B.~S.~DeWitt, 
  Phys.\ Rev. {\bf 160}, 1113 (1967).
\bibitem{mithani:2015PRD}
  A.~T.~Mithani and A.~Vilenkin,
  Phys.\ Rev.\ D {\bf 91}, 123511 (2015).
\bibitem{feng:2017PRD}
  J.~C.~Feng and R.~A.~Matzner,
  Phys.\ Rev.\ D {\bf 96}, 106005 (2017).
\bibitem{albrecht:1982PRL} 
  A.~Albrecht and P.~J.~Steinhardt, 
  Phys.\ Rev.\ Lett {\bf 48}, 1220 (1982).
\bibitem{Jensen:1983ac} 
L.~G.~Jensen and P.~J.~Steinhardt,
Nucl.\ Phys.\ B {\bf 237}, 176 (1984).

\end{thebibliography}
\end{document}